\documentclass[11pt]{article}
\newcommand{\comment} [1]{}
\usepackage[noend]{algorithm,algorithmic}
\usepackage{graphicx,url,times,fullpage,amsmath}
\usepackage{times,amsmath,epsfig}
\usepackage{mathrsfs}
\usepackage{amssymb}
\usepackage{times,amsmath,epsfig}
\usepackage{amsmath,graphicx,latexsym}
\usepackage{subfigure,afterpage}
\usepackage{authblk}

\floatname{algorithm}{Algorithm}

\newtheorem{theorem}{Theorem}

\begin{document}

\title{Privacy Preserving Linear Programming\thanks{The preliminary version appeared in the Proceedings of the 24th ACM Symposium on Applied Computing (SAC '09)}
}

% Put no more than the first THREE authors in the \author command

\author[$\S$]{Yuan Hong}
\author[$\#$]{Jaideep Vaidya}
\author[$\S$]{Nicholas Rizzo}
\author[$\dagger$]{Qi Liu}
\affil[$\S$]{Department of Information Security and Digital Forensics, University at Albany, State University of New York, \{hong, nrizzo\}@albany.edu}
\affil[$\#$]{Department of Management Science and Information Systems, Rutgers University, jsvaidya@cimic.rutgers.edu}
\affil[$\dagger$]{Department of Accounting and Business Law, Siena College, qliu@siena.edu}

\date{}
\maketitle
\begin{abstract}
With the rapid increase in computing, storage and networking resources, data is not only collected and stored, but also analyzed. This creates a serious privacy problem which often inhibits the use of this data. In this chapter, we investigate and resolve the privacy issues in a fundamental optimization problem -- linear programming (LP) which is formulated by data collected from different parties. We first consider the case where the objective function and constraints of the linear programming problem are not partitioned between two parties where one party privately holds the objective function while the other party privately holds the constraints. Second, we present a privacy preserving technique for the case that objective function and constraints are \emph{arbitrarily partitioned} between two parties where each party privately holds a share of objective function and constraints. Finally, we extend the technique for securely solving two-party arbitrarily partitioned linear programming problems to a multi-party scenario. In summary, we propose a set of efficient and secure transformation based techniques that create significant value-added benefits of being independent of the specific algorithms used for solving the linear programming problem.

\end{abstract}

\section{Introduction}\label{sec:intro}
With the rapid increase in computing, storage and networking resources, data is not only collected and stored, but also analyzed. This creates a serious privacy problem which often inhibits the use of this data. In turn, this raises the question of whether it is possible to realize value from distributed data without conflicting with security and privacy concerns? The field of Secure Multiparty Computation (SMC) \cite{Yao86} addresses exactly this problem and provides theoretical foundations for preserving privacy while analyzing data held by distributed sites. The celebrated results \cite{Yao86,Goldreich87,BGW88} in this area show that any function can be securely computed in polynomial time with respect to the size of the circuit required to compute the function. However, this can lead to very inefficient solutions for complex functions or for large-scale input sizes. More efficient privacy preserving solutions are necessary for such important sub-problems.

In this chapter, we look at the specific problem of how organizations optimize the allocation of global resources using linear programming while preserving the privacy of local information. As an important fundamental problem, linear programming (LP) is a subclass of optimization problems where all of the constraints and the objective function are \emph{linear}. It is applicable to a wide variety of real world problems in many industries such as transportation, commodities, airlines, and communication. 

More importantly, many practical linear programming models involve multiple parties to collaboratively formulate and solve the problem with global as well as local information. 
A specific example can be seen in the packaged goods industry, where delivery trucks are empty $25\%$ of the time. Just two years ago, Land O'Lakes truckers spent much of their time shuttling empty trucks down slow-moving highways, wasting several million dollars annually. By using a web based collaborative logistics service (Nistevo.com), to merge loads from different companies (even competitors) bound to the same destination, huge savings were realized via their collaboration (freight costs were cut by $15\%$, for an annual savings of \$2 million\cite{Turban-LoL}). However, this required all the participating parties to send their local data to a centralized site, which charges the collaborative logistics services. Since the disclosed data may contain a considerable amount of proprietary information, different parties' privacy might be compromised in such services. 

Another example is that of Walmart, Target and Costco, who individually, ship millions of dollars worth of goods over the seas every month. These feed into their local ground transportation network. The cost of sending half-empty ships is prohibitive, but the individual corporations have serious problems with disclosing freight information. If they can determine what trucks should make their way to which ports to be loaded onto certain ships i.e. solve the classic transportation problem (modeled as LP problems), without knowing the individual constraints, the savings would be enormous. In all of these cases, complete sharing of data would lead to invaluable savings/benefits. However, since unrestricted data sharing is a competitive impossibility or requires great trust, better solutions must be found.

To tackle the above issues, the proposed technique in this chapter could make such collaboration possible without the release of proprietary information. The rest of this chapter is organized as follows. Section \ref{sec:related} reviews the literature relevant to this study. Section \ref{sec:model} introduces the fundamental LP model and the collaborative LP models, and we also present a transformation-based scheme for a two-party collaborative LP problem in this section. Section \ref{sec:twoparty} presents the algorithms for securely solving two-party collaborative LP problems. Section \ref{sec:mp} extends the algorithm to solve multi-party collaborative LP problems. %Section \ref{sec:exp} demonstrates the experimental results. 
Finally, Section \ref{sec:concl} concludes this chapter and discusses the future work.

\section{Related Work}\label{sec:related}
\subsection{Secure Multiparty Computation}

Privacy preserving linear programming \cite{Vaidya09SAC} is very relevant to a fundamental field -- secure multiparty computation (SMC) \cite{Yao86}. Under the definition of SMC \cite{Yao86}, a computation is secure if at the end of the computation, no party learns anything except its own input and the results. The gold standard of provable security is that a trusted-third party which performs the entire computation would achieve the same result as the SMC algorithm. Yao first postulated the two-party comparison problem (Yao's Millionaire Protocol) and developed a provably secure solution \cite{Yao86}. Goldreich et al. \cite{Goldreich87} generalized this to multiparty computation and proved that there exists a secure solution for any functionality. 

There has been significant theoretical work in this area. Both Yao \cite{Yao86} and Goldreich \cite{Goldreich87} assumed polynomially time bounded passive adversaries. In a line of work initiated by Ben-Or et al. \cite{BGW88}, the computational restrictions on the adversary were removed, but users were assumed to be able to communicate in pairs in perfect secrecy. Ben-Or et al. \cite{BGW88} assume passive adversaries, while Chaum et al. \cite{Chaum88} extend this to active adversaries. Ostrovsky and Yung \cite{OY91Mobile} introduce the notion of mobile adversaries, where the corrupting users may change from round to round. 
Finally, the coercing adversary who can force users to choose their inputs in a way he favors was introduced in the context of electronic elections by Benaloh and Tunistra \cite{BT94Ballot}, and generalized to arbitrary multiparty computation by Canetti and Gennaro \cite{Canetti96incoercible}. Much effort has been devoted to developing crisp definitions of security \cite{Mielikainen04,GL90Def,Canetti92}. However, due to efficiency reasons, it is completely infeasible to directly apply the theoretical work from SMC to form secure protocols for optimization such as linear programming.

\subsection{Collaborative/Distributed Optimization}

Optimization problems occur in all the industries. There is work in distributed optimization (viz. collaborative optimization) that aims to achieve a global objective using only local information. This falls in the general area of distributed decision making with incomplete information. This line of research has been investigated in a worst case setting (with no communication between the distributed agents) by Papadimitriou et al. \cite{Papa91InfoValue,Papa93DistLP}. Papadimitriou and Yannakakis \cite{Papa93DistLP} first explore the problem facing a set of decision-makers who must select values for the variables of a linear program, when only parts of the matrix are available to them and prove lower bounds on the optimality of distributed algorithms having no communication. Awerbuch and Azar \cite{awerbuch94local} proposed a distributed flow control algorithm with a global objective which gives a logarithmic approximation ratio and runs in a polylogarithmic number of rounds. Bartal et al.'s distributed algorithm \cite{Bairs04DistOpt} obtains a better approximation while using the same number of rounds of local communication. Furthermore, some other distributed optimization problems have been studied in literature, such as distributed constraint satisfaction problem (CSP) \cite{Modi03Adopt,Mailler04OptApo} and distributed dynamic programming \cite{Petcu05DPOP,Petcu05distopt}

However, in general, the work in distributed optimization has concentrated on reducing communication costs and has paid little or no attention to security constraints. Thus, some of the summaries may reveal significant information. In particular, the rigor of security proofs has not seen wide spread applications in this area. There is some work in secure optimization. Silaghi and Rajeshirke \cite{Silaghi04} show that a secure combinatorial problem solver must necessarily pick the result randomly among optimal solutions to be really secure. Silaghi and Mitra \cite{SilaghiMitra04} proposed arithmetic circuits for solving constraint optimization problems that are exponential in the number of variables for any constraint graph. A significantly more efficient optimization protocol specialized on generalized Vickrey auctions and based on dynamic programming was proposed by Suzuki and Yokoo \cite{Suzuki04Vickrey}, though it is not completely secure under the framework in \cite{Silaghi04}. Yokoo et al. \cite{Yokoo02DCSP} also proposed a scheme using public key encryption for secure distributed constraint satisfaction. Silaghi et al. \cite{Silaghi06} showed how to construct an arithmetic circuit with the complexity properties of DFS-based variable elimination, and that finds a random optimal solution for any constraint optimization problem. Atallah et al. \cite{sec_supplychain} proposed protocols for secure supply chain management. However, much of this work is still based on generic solutions and not quite ready for practical use. Even so, some of this work can be leveraged to advance the state of the art, by building general transformations or privacy-preserving variants of well-known methods.

\subsubsection{Privacy Preserving Linear Programming}

Recently, there has been significant interest in the area of privacy-preserving linear programming. Work in privacy-preserving linear programming has followed two major directions \cite{HongJIS14}: (1) the problem transformation approach, and (2) the SMC based approach.

\begin{itemize}
\item \textbf{Transformation based approach.} Du \cite{DuThesis} and Vaidya \cite{Vaidya09SAC}
transformed the linear programming problem by multiplying a monomial matrix to both the constraint matrix and the objective function, assuming that one party holds the objective function while the other party holds the constraints. Bednarz et al. \cite{Bednarz2009} pointed out a potential attack to the above transformation approach. To correct the flaw in \cite{Vaidya09SAC}, we revised the transformation and extended the work to the multiparty scenario in this chapter.

In addition, Mangasarian \cite{Mangasarian12,Mangasarian11} presented two transformation approaches for horizontally partitioned linear programs and vertically partitioned linear programs, respectively. Li et al. \cite{LiLD13} extended the transformation approach \cite{Mangasarian12} for horizontally partitioned linear programs with equality constraints to inequality constraints. We have identified a potential inference attack to Mangasarian and Li's transformation based approach, and revised the transformation with significantly enhanced security guarantee in \cite{HongOpt13}. Meanwhile, Hong extended the above transformation to privacy-preserving non-linear programming \cite{HongThesis}, and applied the model to securely anonymize distributed data \cite{HongTDSC15}. Furthermore, Hong et al. \cite{HongDBsec,HongJCS12} proposed approaches to securely solve the arbitrarily partitioned collaborative linear programming problems in both semi-honest and malicious adversarial models. More recently, Dreier and Kerschbaum proposed a secure transformation approach for a complex data partition scenario, and illustrated the effectiveness of their approach \cite{DreierLP}.

\item \textbf{SMC based approach.} Li and Atallah \cite{AtallahPPLP06} addressed the collaborative linear programming problem between two parties where the objective function and constraints can be arbitrarily partitioned, and proposed a secure simplex method for such a problem using cryptographic tools. Vaidya \cite{Vaidya09aina} proposed a secure revised simplex approach also using SMC techniques but improved the efficiency when compared with Li and Atallah 's approach \cite{AtallahPPLP06}. Catrina and Hoogh \cite{CatrinaH10} presented a solution to solve distributed linear program based on secret sharing. The protocols utilized a variant of the simplex algorithm and secure computation with fixed-point rational numbers, optimized for such application.

\end{itemize}

Furthermore, as for securing collaborative combinatorial optimization problems (particularly those NP-hard problems), Sakuma et al. \cite{SakumaK07} proposed a genetic algorithm for securely solving two-party distributed traveling salesman problem (TSP). They consider the case that one party holds the cost vector while the other party holds the tour vector (this is a special case of the multiparty collaborative combinatorial optimization). The TSP that is completely partitioned among multiple parties has been discussed but not solved in \cite{SakumaK07}. Later, Hong et al. \cite{HongDBSEC14} consider a more complicated scenario in TSP where the problem is partitioned to multiple (more than two) parties, and securely solve it with privacy-preserving simulated annealing protocol in \cite{HongDBSEC14}. Meanwhile, Hong et al. \cite{HongTabu} also addressed the privacy concern in another NP-hard problem, graph coloring which is partitioned among two or more parties. Then, the graph coloring problem can be securely solved with the privacy-preserving tabu search protocol.

\section{Models}\label{sec:model}
In this Section, we first review the fundamental LP problem, followed by introducing the LP problems in which the objective function and constraints are partitioned. Furthermore, we propose a transformation-based privacy-preserving approach to solve such LP problems and discuss the potential privacy risks in transformation.

\subsection{Fundamental LP Model}
Optimization is the study of problems in which one seeks to minimize or maximize a real function by systematically choosing the values of real or integer variables from within an allowed set. Formally, given a function $f: A \longrightarrow R$ from some set $A$ to the real numbers, we seek an element $x_0$ in $A$ such that $f(x_0) \leq f(x), \forall x \in A$ (``minimization'') or such that $f(x_0) \geq f(x), \forall x \in A$ (``maximization''). Thus, an optimization problem has three basic ingredients:

\begin{itemize}
\item An \emph{objective function} which we want to minimize or maximize. \vspace{-0.025in}
\item A set of \emph{unknowns} or \emph{variables} which affect the value of the objective function. \vspace{-0.025in}
\item A set of \emph{constraints} that allow the unknowns to take on certain values or exclude others.
\end{itemize}

In linear programming, the objective function $f$ is linear and the set $A$ is specified using only linear equalities and inequalities. Thus, for linear programming, the problem can be easily restated as

\begin{eqnarray*}
\min &c^Tx&\\
\textrm{s.t. } Mx &\geq &B\\
x& \geq& 0
\end{eqnarray*}

\subsection{Collaborative LP Model}

Optimization problems (especially issues in linear programming) have been well studied in the literature -- methods have been proposed for the cases where all of the data is available at a central site. Methods have also been proposed for the cases with incomplete information (distributed optimization). However, all solution methods assume that all the necessary data is centralized or freely available. On the contrary, whenever the data is distributed, we have heightened concerns about problems caused by privacy/security. Different parties might own different constraints or even different parts of the same constraint. Thus, if there are several parties solving a collaborative LP problem, data (constraints or objective function) partitioning should be examined.

\subsubsection{Data Partition}
\label{sec:parti}

There are many ways in which data could be distributed. Each of the ingredients of the optimization problem could be distributed. For example, the objective function might be known to only one party, or parts of it known to some subsets of the parties. The constraints that define the set $A$ might also be distributed in some fashion. Thus, the different ways in which data is distributed give rise to the following categorization.

\begin{itemize}
	\item \textbf{Horizontal Partitioning.} Here, each constraint would be fully owned by one party. Thus, different parties own different constraints. An example of this would be the distributed scheduling problem. Suppose that several schedulers need to schedule tasks on machines. Each task can be executed by several machines (though not all), and it can be split between several machines, but the fraction of all tasks executed by a machine must under no circumstances exceed its capacity. Each scheduler only knows the tasks that may be executed on its pertinent machines, and based on this information it must decide what fractions of its task to send to which machines. The sum of all fractions is to be maximized. Here, the objective function could be known to a single party or to all of the parties, or even be shared by the parties.
	
	\item \textbf{Vertical Partitioning.} In this case, each constraint is shared between some subset of the parties. An example of this would be the organization theory problem. A large enterprise has a very extensive set of tasks -- say, products manufactured. A fundamental question in organization theory is, \emph{how are these tasks to be partitioned among managers?} Although the profitability and resource requirements of these products may change dynamically with market conditions, the \emph{constraint structure}, the sparsity pattern of the constraint matrix of the associated linear program, may be fixed. That is, it is known in advance, which products compete for which resources. What are the \emph{organizational principles} that should guide this assignment of tasks to managers, so that they can make more informed decisions. Again, the objective function might be known to all of the parties, or just to a single party, or be shared by the parties.
	
	\item \textbf{Arbitrary Partitioning.} Apart from the prior two partitioning methods, the data may also be arbitrarily partitioned in some way (some combinations of the above). This is more general and subsumes both of the earlier cases. Completely arbitrary partitioning of data is unlikely in practice, though certain specific configurations might easily be found. In any case, solutions for this case will always work for both of the prior cases as well. Section \ref{sec:arbip} presents the extended secure solution for LP problems with arbitrarily partitioned objective function and constraints with respect to universal cases.
\end{itemize}

\subsubsection{Collaborative LP Problem Discussion}\label{sec:col}
After describing ways of partitioning data, we now discuss how to partition data for some typical collaborative LP problems. In this chapter, we start from a special case of arbitrary partitioning between two parties, where one party owns the objective function while the other party owns the constraints. This may happen, for example, in the case of a manufacturer that would like to evaluate several transportation options between its factories and raw material suppliers, with different transportation options that have different costs (leading to different objective functions). 

After that, we propose privacy preserving solutions for two or multiple parties who own arbitrarily partitioned data by assuming that the global constraints and objectives are the sum of all partitioned data. This happens very often. For example, a winemaking company that grows grapes in various locations, then sends them to different wineries for producing wine. Each vineyard can produce a limited amount of each type of grape; each winery has a requirement for how much of each type of grape it needs. These form the constraints of the LP problem. The goal of the company is to get the grapes to the wineries at the lowest cost (e.g., weight*distance) to get the grapes from the vineyards to the wineries. Obviously, the lowest cost solution would be to have the closest vineyard to each winery produce all the grapes needed by that winery. However, that violates the constraints on how much each vineyard can produce. The goal is to come up with a feasible solution that minimizes cost, where a feasible solution is one that meets all of the constraints imposed by the vineyards. If the winemaking company outsources the transportation to an external company, the collaborative LP problem will turn to the special case, where the external company owns the objective function while the winemaking company owns the constraints. They certainly don't want to reveal their data to the other party.

Now imagine that there are several winemaking companies, where the wineries have their own transportation department and several wineries in some locations. Due to the constraints on how much different vineyards can produce, they want to cooperate with each other on transporting grapes. This is a multiparty collaborative LP problem, where each party owns an objective function and several constraints (if they solve the transport problem independently) while the global objective function and constraints are the sum of all individuals (if they cooperate with each other). In case of this LP problem, each party does not want to reveal anything to any other parties, and each party eventually obtains an individual share of the solution to transport grapes. All the private data (objectives and constraints) are arbitrarily partitioned. In this chapter, we propose privacy-preserving schemes for this general and untrusted collaborative case. Besides the objective function and constraints, the global optimal solution in this case is also the sum of all the solutions of all the parties. The solution for each party means how many grapes should be delivered from each winery location to each vineyard for this party, whereas the global solution represents the total grapes distribution network for all the parties.

Note that the arbitrarily partitioned LP problems are not always formulated as the sum of the private shares (e.g., some parties may share the same constraint in an LP problem). Therefore, in this chapter, we propose privacy-preserving solutions for the LP problems in which the objective and constraints are the sum of the arbitrarily partitioned private shares (e.g., the above winemaking example). We will explore solutions for variant constraints and objectives in our future work.

\subsection{A Transformation Approach for LP Problems}\label{sec:tran}

A possible solution is to transform the vector space by applying a linear transformation. This idea was first proposed by Du to solve systems of linear equations \cite{DuAta012}. Du later extended this idea to solve the two-party linear programming problem \cite{DuThesis}. Later, Vaidya \cite{Vaidya09SAC} proposed a secure transformation based approach for two-party privacy preserving linear programming problems in which one party holds the objective function and the other party holds the constraints, detailed as below. As noted before, assume that you want to solve the problem:

\begin{eqnarray*}
\min &c^Tx&\\
\textrm{s.t. } Mx &\leq &B\\
x& \geq& 0
\end{eqnarray*}

The key to the solution is based on the fact that $MQQ^{-1}x \leq B$, and $Q^{-1}x \geq 0$ if $Mx \leq B$ and $x \geq 0$ where $Q$ should be a monomial matrix (pointed out by Bednarz\cite{Bednarz2009}). Let $M' = M Q$, $y = Q^{-1}x$, and $c'^{T} = c^{T}Q$. We now have a new linear programming problem:

\begin{eqnarray*}
\min &c'^{T}y&\\
\textrm{s.t. } M'y& \leq &B\\
y& \geq& 0
\end{eqnarray*}

Following the methodology of Du \cite{DuThesis}, we can prove that if $y*$ is the solution to the new problem, then $x* = Qy*$ must be the solution to the original problem that minimizes $c^{T}x$. The proof is based on contradiction as follows: Suppose $x* = Qy*$ is not the optimal solution for the original problem. In this case, there must be another vector $x'$ such that $c^Tx' < c^Tx*$ where $Mx' \leq B$ and $x' \geq 0$. Let $y' = Q^{-1}x'$. Now, if we evaluate the objective function of the new problem for this new solution, we find $c'^{T}y' = c^TQQ^{-1}x' = c^Tx'$. Now,

\begin{eqnarray*}
c^Tx' < c^Tx* & \implies &c'^{T}y' < c^Tx*\\
          & \implies &c'^{T}y' < c^TQQ^{-1}x*\\
          & \implies &c'^{T}y' < c'^{T}y*
\parbox[t]{0.6in}{(since~$c'^{T}=c^{T}Q$~and~x*=Qy*)}
\end{eqnarray*}

However, this would imply that $y'$ is a better solution to the new problem than $y*$ which leads to a contradiction (since $y*$ is supposed to be the optimal solution). Thus, it is clear that transformation will give us the correct solution, and we can take use of it to secure the transformation and preserve each party's privacy in the partitioned LP problems.

\section{Secure Two-Party Transformation}\label{sec:twoparty}
 In this section, we first introduce the fundamental encryption method that we are about to use, and then design the two-party secure transformation protocol in two different cases -- on one hand, the constraints matrix and objective function are owned by different parties \cite{Vaidya09SAC}; on the other hand, both of them are split and shared by both parties.

\subsection{Fundamental Encryption}\label{sec:funda}
At the heart, transformations require matrix multiplication, which requires numerous scalar product computations. Thus, multiplying a $m \times n$ matrix with a $n \times n$ matrix simply requires $mn$ scalar products. Goethals et al. \cite{Goethals-scalprod} proposed a simple and provably secure method to compute the scalar product using Homomorphic Encryption, which we now briefly describe. The problem is defined as follows: $P_1$ has an $n$-dimensional vector $\vec{X}$ while $P_2$ has an $n$-dimensional vector $\vec{Y}$. At the end of the protocol, $P_1$ should get $r_a = \vec{X} \cdot \vec{Y} + r_b$ where $r_b$ is a random number chosen from a uniform distribution and is known only to $P_2$. The key idea behind the protocol is to use a homomorphic encryption system such as the Goldwasser-Micali cryptosystem \cite{blum-goldwasser}, the Benaloh cryptosystem \cite{Benaloh87}, the Naccache-Stern cryptosystem\cite{NacStern98}, the Paillier
cryptosystem \cite{Paillier99}, and the Okamoto-Uchiyama cryptosystem \cite{OkaUchi98}. Homomorphic Encryption is a semantically-secure public-key encryption which, in addition to standard guarantees, has the additional property that given any two encrypted messages $E(A)$ and $E(B)$, there exists an encryption $E(A*B)$ such that $E(A)*E(B) = E(A*B)$, where $*$ is either addition or multiplication (in some abelian group). The cryptosystems mentioned above are additively homomorphic (thus the operation $*$ denotes addition). Using such a system, it is quite simple to create a scalar product protocol. The key is to note that $\sum_{i=1}^{n} x_i \cdot y_i = \sum_{i=1}^{n} (x_i + x_i + \dots + x_i)$ ($y_i$ times). If $P_1$ encrypts her vector and sends in encrypted form to $P_2$, $P_2$ can use the additive homomorphic property to compute the dot product. The specific details are given below:

\begin{algorithm}[!h]
\begin{algorithmic}[1]
	
\REQUIRE $P_1$ has input vector $\vec{X} = \{x_1,\dots,x_n\}$
\REQUIRE $P_2$ has input vector $\vec{Y} = \{y_1,\dots,y_n\}$
\ENSURE $P_1$ and $P_2$ get outputs $r_A,r_B$ respectively such that $r_A + r_B = \vec{X} \cdot \vec{Y}$ \STATE $P_1$ generates a private and public key pair (sk, pk). \STATE $P_1$ sends pk to
$P_2$. \FOR{$i = 1 \dots n$}
  \STATE $P_1$ generates a random new string $r_i$.
  \STATE $P_1$ sends to $P_2$ $c_i = Enc_{pk}(x_i;r_i)$.
\ENDFOR \STATE $P_2$ computes $w = \prod_{i=1}^{n} c_i^{y_i}$ \STATE
$P_2$ generates a random plaintext $r_B$ and a random nonce $r$.
\STATE $P_2$ sends to $P_1$ $w' = w \cdot Enc_{pk}(-r_B;r')$. \STATE
$P_1$ computes $r_A = Dec_{sk}(w') = \vec{X} \cdot \vec{Y} - r_B$.
\end{algorithmic}
\caption{Secure scalar product}\label{algo:homo}
\end{algorithm}

\subsection{Two-Party Secure Transformation without Partitioning \cite{Vaidya09SAC}}\label{sec:wopart}
As discussed earlier, the specific problem we look at in this Section consists of a special case of arbitrary partitioning, where one party owns the objective function while the other party owns the constraints. As described in Section \ref{sec:col}, this may happen, for example, in the case of a manufacturer that would like to evaluate a transportation option for shipping its products to different locations, which is provided by an external company. Here, it would know the constraints, while the transporter would know the objective function.

It is quite easy to extend the encryption protocol to work for matrix multiplication and for shared matrices. According to the distributive law of matrix algebra, $M(Q_1 + Q_2) = M Q_1 + M Q_2$ and $C^T(Q_1+Q_2)=C^TQ_1+C^TQ_2$. $P_2$ first computes the encrypted form of $MQ_2$ and sends this to $P_1$ along with the encrypted form of $M$ and $Q_2$. Using the properties of homomorphic encryption, $P_1$ can now compute $MQ_1$ and $C^TQ_1$ in encrypted form, and then compute the encrypted form of $MQ_1 + MQ_2$ and $C^T(Q_1+Q_2)$. $P_1$ then sends them back to $P_2$ who decrypts to finally get $MQ$ and $C^TQ$. The detailed algorithm is given in Algorithm \ref{algo:transform}.

\begin{algorithm}[!h]
\begin{algorithmic}[1]
\REQUIRE $P_1$ has the $n \times n$ matrix $Q_1$ and the
coefficients of the objective function $C^T$ \REQUIRE $P_2$ has the
$n \times n$ matrix $Q_2$ and the $m \times n$ matrix $M$ \STATE
$P_2$ generates a private and public key pair (sk, pk). \STATE $P_2$
computes the $m \times n$ matrix $MQ_2$ \STATE $P_2$ computes
$(MQ_2)' = Enc_{pk}(MQ_2)$ (i.e., the encryption of each element of
$MQ_2$ with pk -- a random nonce is chosen for each encryption)

\STATE $P_2$ computes $Q_2' = Enc_{pk}(Q_2)$ (i.e., the encryption
of each element of $Q_2$ with pk -- a random nonce is chosen for
each encryption)

\STATE $P_2$ computes $M' = Enc_{pk}(M)$ (i.e., the encryption of
each element of $M$ with pk -- again, a random nonce is chosen for
each encryption) \STATE $P_2$ sends pk, $M'$, $Q_2'$ and $(MQ_2)'$
to $P_1$

\COMMENT{$P_1$ now computes the encrypted matrix $S'$ and vector
$V'$ as follows}

\FOR{each row $i$ of $M'$ and each column $j$ of $Q_1$}
  \STATE $P_1$ computes $S'_{ij} = \prod_{k=1}^{n} M'^{{Q_1}_{kj}}_{ik} *
(MQ_2)'_{ij}$ \ENDFOR

\STATE $P_1$ computes $(C^T Q_1)' = Enc_{pk}(C^T Q_1)$ (i.e., the
encryption of each element of $C^T Q_1$ with pk -- a random nonce is
chosen for each encryption)

\FOR{each each column $i$ of $Q_2'$}
  \STATE $P_1$ computes $V'_{i} = \prod_{k=1}^{n} Q_2'^{C^T_k} *
(C^T Q_1)'_{i}$ \ENDFOR

\STATE $P_1$ sends $S'$ and $V'$ to $P_2$ \STATE $P_2$ computes $S =
Dec_{sk}(S') = M (Q_1 + Q_2)$ and $V=C^T(Q_1+Q_2)$
\end{algorithmic}
\caption{Secure two-party transformation without partitioning}\label{algo:transform}
\end{algorithm}

$P_1$ owns the objective function $C^TX$ and its transformation matrix $Q_1$, whereas $P_2$ holds the constraints matrix $M$ and its transformation matrix $Q_2$. We can analyze the
security/privacy of Algorithm \ref{algo:transform} by looking at the data held, received, computed and decrypted by each party, as shown in Table \ref{table:data1}.

\begin{table}[!h]
   \centering
      \begin{tabular}{|c|c|c|}
      \hline
      \ &$P_1$&$P_2$\\
      \hline
      Hold&$Q_1, C^T$&(sk, pk), $Q_2, M$\\
      \hline
      Receive&$(MQ_2)', Q_2', M'$, pk&$S', V'$\\
      \hline
      Compute&$S', (C^TQ_1)', V'$&$(MQ_2)', Q_2', M'$\\
      \hline
      Decrypt&&$M(Q_1+Q_2), C^T(Q_1+Q_2)$\\
      \hline
      \end{tabular}

\caption{Data in Algorithm \ref{algo:transform}}\label{table:data1}
\end{table}

\begin{enumerate}
\item $P_1$ holds $Q_1$ and $C^T$ while $P_2$ holds $Q_2$ and $M$. All of them are private components for themselves.
\item $P_1$ receives $(MQ_2)'$, $Q_2'$ and $M'$ from $P_2$, but $P_1$ cannot decrypt them with a public key pk; $P_1$ computes(encrypts) $S'$, $(C^TQ_1)'$ and $V'$ using pk, but nothing can be learned from this step.
\item $P_2$ decrypts $S'$ and $V'$ and obtains $M(Q_1+Q_2)$ and $C^T(Q_1+Q_2)$. Even if $P_2$ owns $M$, it cannot compute $Q=Q_1+Q_2$ and $C^T$.
\end{enumerate}

Finally, $P_2$ solves the new LP problem and shares the new solution with $P_1$, but $P_2$ still does not know the original solution of $P_1$. Hence, Algorithm \ref{algo:transform} is secure.

\subsection{Two-Party Secure Transformation with Arbitrary Partitioning}\label{sec:arbip}

In case that both the objective function $C^T$ and the constraint matrix $M$ of an LP problem are arbitrarily partitioned to two shares, where $C^T=C_1^T+C_2^T$ and $M=M_1+M_2$ (note that $C_1^T$, $C_2^T$ and $C^T$ have the same size; $M_1$, $M_2$ and $M$ have the same size). To securely solve such problem, we extend Algorithm \ref{algo:transform} to Algorithm \ref{algo:transformarb}.

\begin{algorithm}[!h]
\begin{algorithmic}[1]
\REQUIRE $P_1$ has the $n \times n$ matrix $Q_1$, the share of objective function $C_1^TX$ and the $m \times n$ matrix $M_1$
\REQUIRE $P_2$ has the $n \times n$ matrix $Q_2$, the share of objective function $C_2^TX$ and the $m \times n$ matrix $M_2$
\\
\COMMENT{All the encryptions are based on each elements in the matrices and a random nonce is chosen for each encryption} 

\STATE $P_1$ generates a private and public key pair ($sk$, $pk$) 

\STATE $P_1$ computes the $m \times n$ matrix: $M_1' = Enc_{pk}(M_1)$ and the $n$-dimensional
vector: $C_1'=Enc_{pk}(C_1)$.

\STATE $P_1$ sends $pk$, $M_1'$ and $C_1'$ to $P_2$

\STATE $P_2$ computes $M_2' = Enc_{pk}(M_2)$, $C_2'=Enc_{pk}(C_2)$, $M' = M_1'*M_2' =
Enc_{pk}(M_1+M_2)$, $C'=Enc_{pk}(C_1)*Enc_{pk}(C_2)$ and $Q_2'=Enc_{pk}(Q_2)$. ($*$ represents the multiplication operation for each element in the same position of two matrices)

\FOR{each row $i$ of $M'$ and each column $j$ of $Q_2$}
  \STATE $P_2$ computes $(MQ_2)'_{ij} = \prod_{k=1}^{n} (M_{ik})'^{{Q_2}_{kj}}$
\ENDFOR

\FOR{each row $i$ of $C'$ and each column $j$ of $Q_2$}
\STATE $P_2$ computes $(C^TQ_2)'_{ij} = \prod_{k=1}^{n} (C_{ik})'^{{Q_2}_{kj}}$
\ENDFOR

\STATE $P_2$ sends $(MQ_2)'$ and $(C^TQ_2)'$ back to $P_1$

\FOR{each row $i$ of $(MQ_2)'$ and each column $j$ of $Q_1$}
\STATE $P_1$ computes $(MQ_2Q_1)'_{ij} = \prod_{k=1}^{n} ((MQ_2)_{ik})'^{{Q_1}_{kj}}$
\ENDFOR

\FOR{each row $i$ of $(C^TQ_2)'$ and each column $j$ of $Q_1$}
\STATE $P_1$ computes $(C^TQ_2Q_1)'_{ij} = \prod_{k=1}^{n} ((C^TQ_2)_{ik})'^{{Q_1}_{kj}}$
\ENDFOR

\STATE $P_1$ decrypts $(MQ_2Q_1)'$ and $(C^TQ_2Q_1)'$ with its private key $sk$
\end{algorithmic}
\caption{Secure two-party transformation with arbitrary partitioning}\label{algo:transformarb}
\end{algorithm}

Without loss of generality, in Algorithm \ref{algo:transformarb}, $P_1$ solves the LP problem and it eventually obtains $C^TQ$ and $MQ$ after decryption, where $Q=Q_1Q_2$. Specifically, $P_1$ first sends its transformed and encrypted matrix/vector and the public key to $P_2$, and receives the encrypted and transformed overall constraints matrix from $P_2$. Then, $P_1$ computes the ciphertext of the transformed matrix/vector. Finally, it decrypts the transformed matrix $(MQ_2Q_1)'$ and vector $(C^TQ_2Q_1)'$ as well as solve the transformed LP problem.

After solving the transformed problem to get the optimal solution $y*$, $P_1$ and $P_2$ jointly reconstruct $x*=Q_2Q_1y*$. Bednarz et al. \cite{Bednarz2009} proposed a possible attack on inferring $Q$ with known $C^TQ, C^T, y*$ and $x*=Qy*$: first, since $Q$ is known as a monomial matrix, all the permutations of $Q$ can be enumerated and computed according to the known vectors $C^TQ$ and $C^T$; second, the adversary can determine the exact case of $Q$ in all the permutations by verifying $x*=Qy*$ one by one.  However, the attack can be eliminated in our secure transformation for arbitrarily partitioned LP problems. First, $x*$ is partitioned and each share of the solution is kept private, thus no solution can be used to verify the permutations. Second, $C^T$ is partitioned and each share of $C^T$ can be permuted before transformation as well. Since $C^T$ is unknown to $P_2$, $P_2$ cannot compute $Q$ based on an unknown vector $C^T$ and the transformed objective vector $C^TQ$. We discuss the security of Algorithm \ref{algo:transformarb} in Section \ref{twosec}.

Furthermore, without loss of generality, we let $P_1$ solve the LP problem. Since either $P_1$ or $P_2$ can be the party to solve the transformed problem (in fact an untrusted third party can also be used to do the same), Algorithm \ref{algo:transformarb} does not violate the fairness in the secure computation.

\subsubsection{Security Analysis}
\label{twosec}

\begin{theorem}
	$P_1$ learns only $MQ_2Q_1$ and $C^TQ_2Q_1$, and $P_2$ learns nothing in Algorithm \ref{algo:transformarb}.
\end{theorem}

\begin{table}[!h]
	\centering
	\begin{tabular}{|c|c|c|}
		\hline
		\ &$P_1$&$P_2$\\
		\hline
		Hold&$Q_1$, (sk, pk), $M_1, C_1$&$Q_2, M_2, C_2$\\
		\hline
		Receive&$(MQ_2Q_1)'$ and $(C^TQ_2Q_1)'$&$M_1', pk$\\
		\hline
		Encrypt or Compute&$(MQ_2Q_1)'$, $(C^TQ_2Q_1)'$, $M_1'$, $C_1'$&$C', M', (MQ_2)'$\\
		\hline
		Decrypt&$MQ_2Q_1$ and $C^TQ_2Q_1$&\\
		\hline
	\end{tabular}
	
	\caption{Each Party's Messages in Algorithm
		\ref{algo:transformarb}}\label{table:data2}
\end{table}

\noindent\textbf{Proof.} Before showing what $P_1$ and $P_2$ can learn in each step of Algorithm \ref{algo:transformarb}, we first look at different messages (e.g., matrices, vectors) each party can obtain in Table \ref{table:data2}. We thus
have:

\begin{enumerate}
\item $P_1$ holds $M_1$ and $Q_1$ while $P_2$ holds $M_2$, and $Q_2$. All of them are private data for themselves;
\item $P_2$ acquires $M_1′$ from $P_1$, but $P_2$ cannot decrypt them with a public key pk;
\item $P_2$ computes(encrypts) $C'$, $M′$, and $(MQ_2)′$ using pk, but nothing can be learned from these steps, since all the operations are based on public key (pk) encrypted objects;

\item $P_1$ decrypts $(MQ_2Q_1)′$ and $(C^TQ_2Q_1)′$ to get $MQ_2Q_1$ and $C^TQ_2Q_1$.
\end{enumerate}

In Algorithm \ref{algo:transformarb}, on one hand, $P_1$ can only obtain $(M_1+M_2)Q_2Q_1$ and $(C_1+C_2)^TQ_2Q_1$. It's impossible to calculate $M_2$, $C_2$ and $Q_2$ with known matrices $M_1$, $Q_1$ and vector $C_1$. Also, the probability of guessing such private numbers is fairly low (note that in the earlier discussion, the adversary does not have any disclosed solutions and objective vectors to verify and determine $Q$ using Bednarz et al. \cite{Bednarz2009}'s attack scenario). Specifically, even if $P_1$ can obtain the $m\times n$ matrix $(M_1+M_2)\times Q_2$ and $n$-dimensional vector $(C_1+C_2)^TQ_2$ from $P_2$ by decrypting the ciphertexts obtained from $P_2$, it is secure because $P_1$ does not
know any factor of the product of matrix/vector multiplication (none of M, C and $Q_2$ can be inferred). Thus, $P_1$ cannot obtain the private data (the share of constraint matrix/objective and the transformation matrix) from the other party $P_2$. On the other hand, with a public key encryption for $M_1$, it is also impossible for $P_2$ to discover $M_1$ or $Q_1$ from the given matrices. Therefore, Algorithm \ref{algo:transformarb} is secure. $\hfill\boxdot$

\section{Secure Multi-Party Transformation}\label{sec:mp}
While Algorithm \ref{algo:transformarb} is limited to two parties, in general any cooperative computation is likely to involve more than two parties (i.e. more than two corporations cooperate to ship goods). Therefore, in this section, we extend the algorithm to  multiple parties (more than two) with arbitrary partitioning.

\subsection{Securing Multiparty Transformation}
\label{sec:securemp}

In multi-party arbitrarily partitioned LP, we can locally encrypt the matrix transformation and multiplication for all parties using their own public keys, and each party decrypts its transformed matrices which are encrypted, transformed and transferred by other parties. 

Specifically, to enforce privacy protection, each party is allowed to generate and send random matrices/vectors for each encryption. Suppose that we are securing an $l$-party LP problem, each party $P_i$ first distributes its public key $pk_i$ to all the other parties. Then, $P_i$ encrypts and transforms its
matrix share $M_i$ with its privately held monomial matrix $Q_i$ and public key $pk_i$: $Enc_{pki} (M_iQ_i)$. Later, all parties jointly transform party $P_i$'s matrix $M_i$ to obtain $Enc_{pki}(M_iQ_1\cdots Q_l + R)$ where $R$ is an encrypted random matrix co-held by all parties. Finally, each party $P_i$ decrypts its transformed matrix with its private key $sk_i$, and all the parties securely sum the transformed matrix $MQ = (M_1+· · ·+M_l)Q_1\cdots Q_l$.  Table \ref{table:multi} shows each party’s matrices after decrypting the summed matrices. Some essential points are worth noting as below:

\begin{itemize}
	\item $*$ stands for the multiplication on each element in the same position of two matrices rather than matrix
	multiplication;
	
	\item $R_{(i,j)}$ stands for the random matrix generated by party $P_j$ for transformation $P_i$'s matrix share $M_i$;
	
	\item If substituting $M_i$ with $C^T_i$, the table will be each party’s objector vectors.
\end{itemize}

Then, Algorithm \ref{algo:transformMulti1} presents the details of the secure multiparty transformation.

\begin{table}[!h]
	\centering
	\begin{tabular}{|c|c|c|c|}
		\hline
		\ $P_1$&$P_2$&...&$P_l$\\
		\hline
		$M_1Q_1Q_2\cdots Q_l+\sum_{j=2}^lR_{(1,j)}$&$-R_{(1,2)}$&...&$-R_{1,l}$\\

		\hline
		$-R_{(2,1)}$&$M_2Q_1Q_2\cdots Q_l+\sum_{j=1, j\ne 2}^lR_{(2,j)}$&...&$-R_{(2,l)}$\\
		\hline
		...&...&...&...\\
		\hline
		$-R_{(l,1)}$&$-R_{(l,2)}$&...&$M_1Q_1Q_2\cdots Q_l+\sum_{j=1}^{l-1}R_{(l,j)}$\\
		\hline
	\end{tabular}
	
	\caption{Each Party's Matrices in a Multi-Party LP (Arbitrary Partitioning) after Decryption}\label{table:multi}
\end{table}

\begin{algorithm}[!h]
\begin{algorithmic}[1]
\REQUIRE $P_i$ has the $n \times n$ matrix $Q_i$, $m \times n$ matrix $M_i$ and $n$-dimensional vector $C_i$ ($1\leq i\leq l$);

\FOR{$ith$ party ($i=1,2,3,\dots,l$)}

\STATE $P_i$ generates a key pair ($sk_i$, $pk_i$) and sends $pk_i$ to all the remaining parties

\STATE $P_i$ computes $(M_iQ)'=Enc_{pk_i}(M_i)$ and $(C_i^TQ)'=Enc_{pk_i}(C_i^TQ_i)$ 

\STATE $P_i$ sends $(M_iQ)'$ and $(C_i^TQ)'$ to the next party

\FOR{$jth$ party $P_j (j=1,2,\dots, l$ and $j\ne i)$}

\STATE $P_j$ generates an $m\times n$ random matrix $R_{(i,j)}$ and encrypts it with $pk_i: R_{(i,j)}'=Enc_{pk_i}(R_{(i,j)})$

\STATE $P_j$ generates an $n$-dimensional random matrix $S_{(i,j)}$ and encrypts it with $pk_i: S_{(i,j)}'=Enc_{pk_i}(S_{(i,j)})$

\FOR{each row $a$ of $(M_iQ)'$ and each column $b$ of $Q_j$}

\STATE $P_j$ computes $(M_iQ)_{(i,j)_{ab}}'=\prod_{h=1}^{n}M_iQ_{ah}'^{({Q_j})_{hb}}*{R_{(i,j)}}_{ab}'$

\ENDFOR

\FOR{each entry $a$ of $(C_i^TQ)'$ and each column $b$ of $Q_j$}

\STATE $P_j$ computes $(C_i^Q)_{(i,j)_{ab}}'=\prod_{h=1}^{n}C_i^TQ_{ah}'^{({Q_j})_{hb}}*{S_{(i,j)}}_{ab}'$

\ENDFOR

\STATE $P_j$ sends the updated $(M_iQ)'$ and $(C^TQ)'$ to the next party

\ENDFOR

\STATE Finally, the last party sends $(M_iQ)'$ and $(C^TQ)'$ back to party $P_i$

\ENDFOR

\FOR{$ith$ party ($i=1,2,3,\dots,l$)} 

\STATE $P_i$ decrypts $(M_iQ)'$ and $(C^TQ)'$ with its private key $sk_i$ to obtain $M_iQ_1Q_2\cdots Q_l+\sum_{i=1,i\ne j}^{l}R_{(i,j)}$ and $C_i^TQ_1Q_2\cdots Q_l+\sum_{i=1,i\ne j}^{l}S_{(i,j)}$

\STATE $P_i$ vertically subtracts all the random matrices/vectors (generated by itself), and obtains $M_iQ_1Q_2\cdots Q_l+\sum_{i=1,i\ne j}^{l}R_{(i,j)}-\sum_{j=1,j\ne i}^lR_{(j,i)}$, and $C_i^TQ_1Q_2\cdots Q_l+\sum_{i=1,i\ne j}^{l}S_{(i,j)}-\sum_{j=1,j\ne i}^lS_{(j,i)}$

\STATE All the parties sum the matrices to obtain: $MQ=(M_1+\dots+M_l)Q_1\cdots Q_l$ and $C^TQ=(C_1+\dots+C_l)^TQ_1\cdots Q_l$

\ENDFOR

\end{algorithmic}
\caption{Secure multi-party transformation with arbitrary partitioning}\label{algo:transformMulti1}
\end{algorithm}

\subsection{Security Analysis}\label{sec:secmp}

We have analyzed the security of Algorithm \ref{algo:transformarb} in Section 5. Similarly, we can analyze all the messages in Algorithm \ref{algo:transformMulti1}.

\begin{theorem}
	Algorithm \ref{algo:transformMulti1} ensures: $\forall i\in [1,l], P_i$ cannot learn $\forall j\in[1,l]$ and $j\ne i$, $M_j$ and $Q_j$.
\end{theorem}

\noindent\textbf{Proof.} Table \ref{table:data3} presents all the exchanged messages in Algorithm \ref{algo:transformMulti1}:

\begin{table}[!h]
	\centering
	\begin{tabular}{|c|c|c|}
		\hline
		 &$P_i(i=1,2,...,l)$&$P_j(j\neq i, j=1,2,...,l)$\\
		
		\hline
		Hold&$Q_i, (pk_i, sk_i), M_i, C_i^T$&$Q_j, (pk_j, sk_j), M_j, C_j^T$\\
		\hline
		Receive&$pk_j, Enc_{pk_i}(M_iQ_1\cdots Q_l)*\prod_{\forall j\ne i}*R_{(i,j)}'$,& $pk_i, Enc_{pk_j}(M_jQ_1\cdots Q_l)*\prod_{\forall i\ne j}*R_{(j,i)}'$, \\
		&$Enc_{pk_j}(M_jQ_jQ_1\cdots Q_i)*R_{(j,1)}'***$&$Enc_{pk_i}(M_iQ_iQ_1\cdots Q_i)*R_{(i,1)}'***$\\
		\hline
		Encrypt&$Enc_{pk_j}(M_jQ_jQ_1\cdots Q_i)*R_{(j,1)}'***R_{(j,i)}'$&$Enc_{pk_i}(M_iQ_iQ_1\cdots Q_i)*R_{(i,1)}'***R_{(i,j)}'$\\
		\hline
		Decrypt &$M_iQ_1\cdots Q_l+\sum_{j=1,j\ne i}^{l}R_{(i,j)}$& $M_jQ_1\cdots Q_l+\sum_{i=1,i\ne j}^{l}R_{(j,i)}$\\
		\hline
		Sum &the sum of $M_iQ$ and random matrices&the sum of $M_jQ$ and random matrices\\
		\hline
	\end{tabular}
	
	\caption{Each Party's Messages in Algorithm \ref{algo:transformMulti1}}\label{table:data3}
\end{table}

We consider two different categories of collaborative parties: one is an arbitrary party $P_i$ while the other one is another arbitrary party $P_j$. We can learn from Algorithm \ref{algo:transformMulti1}, Table \ref{table:data2}, and \ref{table:data3} as below:

\begin{enumerate}
\item $P_i$ receives $\forall j=1,2,3,...,l$ and $j\neq i$, $pk_j$, $Enc_{pk_i}(M_iQ_1\cdots Q_l)*\prod*R_{(i,j)}'$ and $Enc_{pk_j}(M_jQ_jQ_1\cdots)*R_{(j,1)}'***$ from other parties;

\item $P_i$ can neither decrypt $Enc_{pk_j}(M_jQ_jQ_1\cdots)*R_{(j,1)}'***$ nor compute the random matrices and other parties' transformation matrices $Q_j$ from its decrypted matrix $M_iQ_1\cdots Q_l+\sum_{j=1,j\ne i}^{l}R_{(i,j)}$;

\item If substituting the constraint matrix share with the objective vector share, the communication remains secure for all the parties;

\item Algorithm \ref{algo:transformMulti1} does not suffer from Bednarz et al. \cite{Bednarz2009}'s attack for the same reason as Algorithm \ref{algo:transformarb};

\item If more parties are involved in this multiparty LP problem, the protocol is more secure due to an increasing number $l$ which further minimizes the probabilities of guessing the correct matrices and vectors;

\item $l>2$ in Algorithm \ref{algo:transformMulti1}. 
\end{enumerate}

Finally, each party provides a vertical sum of all the matrices and vectors in Table \ref{table:multi}. This is secure to all the parties and no one can learn any other party's private share of the constraints, vectors or transformation matrix from the sum and the data they provide to other parties. $\hfill\boxdot$

Similar to the two-party arbitrarily partitioned LP problem, we can let an arbitrary party solve the LP problem. After solving the transformed problem (optimal solution $y*$), all the parties can jointly reconstruct the original optimal solution $\prod_{i=1}^lQ_iy*$.

\section{Concluding Remarks}\label{sec:concl}
This chapter presents the privacy-preserving linear programming problem, and motivates several different data distributions possible in real life LP models. We present a practical transformation based solution that can solve the special two-party case where one party owns the objective while the other owns the constraints. We also extends the above approach to tackle such issues in a more general case that arbitrary partitioned constraints and objective function are shared by both parties. In addition to a two-party secure computation, we propose an approach to secure multi-party transformation on their matrices and vectors in distributed LP problems. In all our secure algorithms, each party owns a part of the collaborative components as its privacy, eventually acquires the collaborative LP solution, calculates the individual solution using a reverse transformation from the common solution and implements it during the cooperation. We consider the individual transformation matrix, share of the objective vectors and constraints matrices as the private information, and take advantage of our algorithm to preserve them in the transformation and communication.

However, several questions remain for the future. First, the security implied by the solution is still somewhat heuristic. While it is clear that the original constraints, objective function and transformation matrix cannot be reconstituted or calculated from the data each party gets from the others, what about other features. Can anything be inferred about the hardness of the problem, or the type of constraints, or their relation to each other? All of these are significant questions that need to be looked at in more detail for practical deployment of the solution. Some of the work in secure computation shows possible approaches. If we could show that the results of any function computed from the constraints in polynomial time are indistinguishable from random numbers uniformly generated, we can prove that transformation leaks nothing. This needs to be further explored. Another observation is that it should be possible to adapt the transformation approach to work for integer programming as well as quadratic programming. For integer programming the case is the same as the case as for linear programming. However, for quadratic programming, the form of the constraints changes -- therefore the transformation method must take this into account. Nevertheless, polynomial evaluation may still be used to perform the transformation. We intend to further explore all of these problems in the future. Finally, as the intersection of two fundamental fields: secure computation and optimization, privacy preserving linear programming can be applied to tackle the privacy concerns in many real world applications or systems, such as web search \cite{HongCikm09,HongEDBT12,HongWi11}, data mining \cite{LuVAH09,HongTDSC12,HongICDMW12,HongWI13,LuHYFD14,LuHYDB15}, transportation \cite{HongICCVE15}, smart grid \cite{HongIJER15}, and healthcare systems \cite{liu2008securing,hong2008preserving,lu2007access,hong2007hierarchical}. We will explore such real world applications in our future work.

\section*{Acknowledgments}

This work is partially supported by the National Science Foundation
under Grants No. CNS-0746943 and CNS-1618221.

\end{document}